\documentclass{article}

\usepackage{arxiv}
\usepackage[utf8]{inputenc} 
\usepackage[T1]{fontenc}    
\usepackage{url}            
\usepackage{nicefrac}       
\usepackage{float}
\usepackage{microtype}
\usepackage{graphicx}
\usepackage{booktabs} 
\usepackage{lipsum}
\usepackage{xspace}
\usepackage{enumitem}
\usepackage{natbib}
\usepackage[dvipsnames]{xcolor}

\usepackage[breaklinks=true,colorlinks,bookmarks=false,citecolor=Blue,linkcolor=blue]{hyperref}

\usepackage{enumitem}
\usepackage{doi}

\usepackage{amsmath}
\usepackage{amssymb}
\usepackage{mathtools}
\usepackage{amsthm}

\usepackage[capitalize,noabbrev]{cleveref}

\usepackage{fourier}

\usepackage[textsize=tiny]{todonotes}
\usepackage[most]{tcolorbox}
\usepackage{xurl}

\usepackage{tabularx}

\title{AI Risk Management Should Incorporate\\Both Safety and Security}

\newcommand\blfootnote[1]{
  \begingroup
\renewcommand\thefootnote{}\footnote{#1}%
  \addtocounter{footnote}{-1}%
  \endgroup
}

\newcommand*{\affmark}[1][*]{\textsuperscript{\textnormal{#1}}}

\author{
\textbf{Xiangyu Qi}\affmark[1], \textbf{Yangsibo Huang}\affmark[1], \textbf{Yi Zeng}\affmark[2], \textbf{Edoardo Debenedetti}\affmark[3], \textbf{Jonas Geiping}\affmark[4,5], \textbf{Luxi He}\affmark[1], \textbf{Kaixuan Huang}\affmark[1], \and \textbf{Udari Madhushani Sehwag}\affmark[6,7], \textbf{Vikash Sehwag}\affmark[8], \textbf{Weijia Shi}\affmark[9], \textbf{Boyi Wei}\affmark[1], \textbf{Tinghao Xie}\affmark[1], \textbf{Danqi Chen}\affmark[1], \textbf{Pin-Yu Chen}\affmark[10],\and \textbf{Jeffrey Ding}\affmark[11], \textbf{Ruoxi Jia}\affmark[2], \textbf{Jiaqi Ma}\affmark[12], \textbf{Arvind Narayanan}\affmark[1], \textbf{Weijie J. Su}\affmark[13], \textbf{Mengdi Wang}\affmark[1], \textbf{Chaowei Xiao}\affmark[14,15],\and \textbf{Bo Li}\affmark[12,16],  \textbf{Dawn Song}\affmark[17], \textbf{Peter Henderson}\affmark[1], \textbf{Prateek Mittal}\affmark[1]\\
~\\
\affmark[1]Princeton University~~~\affmark[2]Virginia Tech~~~\affmark[3]ETH Zurich~~~\affmark[4]ELLIS Institute Tübingen~~~\affmark[5]MPI for Intelligent Systems\\\affmark[6]Stanford University~~~~\affmark[7]{JPMorgan AI Research}~~~~\affmark[8]Sony AI~~~~\affmark[9]University of Washington~~~~\affmark[10]IBM Research\\\affmark[11]George Washington University~~~~\affmark[12]University of Illinois at Urbana-Champaign~~~~\affmark[13]University of Pennsylvania\\\affmark[14]University of Wisconsin, Madison~~~~\affmark[15]NVIDIA~~~~\affmark[16]University of Chicago~~~~\affmark[17]UC Berkeley
}

\renewcommand{\shorttitle}{AI Risk Management Should Incorporate Both Safety and Security}

\usepackage{pythonhighlight}
\usepackage{booktabs}
\usepackage{graphicx}
\usepackage{booktabs}
\usepackage{caption}
\usepackage{subcaption}
\usepackage{fourier}
\usepackage{multirow}
\usepackage{comment}
\usepackage{wrapfig}

\usepackage{enumitem}
\usepackage{array}

\newcolumntype{C}[1]{>{\centering\arraybackslash}m{#1}}
\newcolumntype{L}[1]{>{\raggedright\arraybackslash}m{#1}}

\usepackage{color}

\definecolor{deepred}{rgb}{0.631,0.102,0.102}
\definecolor{amethyst}{rgb}{0.6, 0.4, 0.8}
\definecolor{darkgreen}{rgb}{0.3,0.7,0.3}
\definecolor{salmon}{RGB}{241, 150, 141}

\newcommand{\xiangyu}[1]{\textcolor{red}{Xiangyu: {#1}}}

\newcommand{\yi}[1]{\textcolor{deepred}{Yi: {#1}}}

\newcommand{\lucy}[1]{\textcolor{amethyst}{Lucy: {#1}}}

\begin{document}
\maketitle
\blfootnote{Correspondence to: Xiangyu Qi~(\url{xiangyuqi@princeton.edu}), Yangsibo Huang~(\url{yangsibo@princeton.edu}),\\Peter Henderson~(\url{peter.henderson@princeton.edu}), Prateek Mittal~(\url{pmittal@princeton.edu})}

\begin{abstract}
The exposure of \textbf{security} vulnerabilities in \textbf{safety}-aligned language models, e.g., susceptibility to adversarial attacks, has shed light on the intricate interplay between AI safety and AI security. Although the two disciplines now come together under the overarching goal of AI risk management, they have historically evolved separately, giving rise to differing perspectives. Therefore, in this paper, we advocate that stakeholders in AI risk management should be aware of the nuances, synergies, and interplay
between safety and security, and unambiguously take into account the perspectives of both disciplines in order to devise mostly effective and holistic risk mitigation approaches. Unfortunately, this vision is often obfuscated, as the definitions of the basic
concepts of "safety" and "security" themselves are often inconsistent and lack consensus across communities. With AI risk management being increasingly cross-disciplinary, this issue is particularly salient. In light of this conceptual challenge, we introduce a unified reference framework to clarify the differences and interplay between AI safety and AI security, aiming to facilitate a shared understanding and effective collaboration across communities.
\end{abstract}

\section{Introduction}

The rapid advancements in artificial intelligence (AI), particularly in the development of large language models (LLMs) \citep{brown2020language,bommasani2021opportunities,chatgpt,gpt4v,openai2023gpt4,touvron2023llama,touvron2023llama-2,claude,geminiteam2023gemini}, have sparked widespread concerns about the potential risks associated with their deployment. As these models become increasingly capable and are targeted for a wide range of tasks, the stakes of potential misuse and the range of possible harm in cases of AI failures grow higher.

In response to these concerns, government bodies are actively considering standards and policies for transparency, monitoring, licensing, and liability of AI technologies \citep{eu-ai-act,EOWhiteHouse,uk-ai-safety-summit}. These efforts have been accompanied by the formation of new government bodies, such as the National Institute of Standards and Technology (NIST)\footnote{\url{https://www.nist.gov/artificial-intelligence/artificial-intelligence-safety-institute}} and United Kingdom (UK) AI Safety Institutes.\footnote{\url{https://www.gov.uk/government/publications/ai-safety-institute-overview/introducing-the-ai-safety-institute}} However, there has been widespread disagreement on the concrete scope of these AI risk management efforts, partially due to the terminology used and the wide range of communities focused on reducing risks from different perspectives.
The scoping of the NIST AI Safety Institute, for example, was hotly debated in a panel convened by the National AI Advisory Committee, where some expressed concern about expanded definitions of AI Safety, while others pushed back, noting that meeting safety objectives requires an expanded scoping~\citep{AlbergottiSemafor2024}.

In this work, we focus on another challenge: potential ambiguities on whether security objectives should be included in the scope of work.
Historically, safety and security objectives have been distinguished across a broad range of established sectors, such as aviation, nuclear energy, chemistry, power grids, and information systems \citep{anderson2022safety-security,pietre2010sema}. For example, in the nuclear domain, safety and security were considered separately in negotiations and treaties, with safety focusing on preventing accidents and unintended harm, while security dealt with preventing malicious actors from causing harm \citep{ding2024nuclear,IAEA2023}. As such, information sharing, legal frameworks, and policies have evolved differently for nuclear safety and nuclear security, with the \citet{IAEA2023} recently suggesting that nuclear safety and nuclear security communities would benefit from increased collaboration and knowledge sharing. 

Even in the context of AI, recent grantmaking efforts by the NSF reflect a distinction between safety and security objectives. The  \textit{Safe Learning-Enabled Systems} program funded by the U.S. National Science Foundation explicitly excludes \textit{"research on securing learning-enabled systems against
adversaries"} out of its scope\footnote{\url{https://www.nsf.gov/pubs/2023/nsf23562/nsf23562.pdf}}. Meanwhile, the \textit{Guaranteeing AI Robustness against Deception (GARD) program} created by the U.S. Defense Advanced Research Projects Agency (DARPA) has been supporting the development of AI security solutions to defend against adversarial attacks on AI systems\footnote{\url{https://www.darpa.mil/news-events/2019-02-06}}. 
The AI Risk Management Framework~\citep{nist_ai_risk} developed by the U.S. National Institute of Standards and Technology~(NIST) defines safety and security as two separate characteristics of trustworthy AI. It suggests that AI safety should take cues from and align with efforts, guidelines, and standards for safety in other established sectors, such as transportation and healthcare; while NIST Cybersecurity Framework~\citep{nist_security_risk} is recommended as a reference for AI security. This further invites more diverse opinions from all those related sectors.

In AI academic research, AI safety and AI security have also developed somewhat separately, with more emphasis on different objectives. For example, \citet{amodei2016concrete} frame the discussion of AI safety more conservatively, with a bias towards the problem of accidents in machine learning (ML) systems, while acknowledging security as a closely related field. But \citet{hendrycks2021unsolved} more directly position many AI security problems under the considerations of AI safety. Interestingly, there are also definitions from the security community that consider AI security as a superset of AI safety instead~\citep{joseph2023terminology}. 
\citet{khlaaf2023toward} would disambiguate between alignment, systems safety, and security as distinct and nuanced concepts.
And others have argued for an expanded sociotechnical framing of AI Safety~\citep{raji2023concrete}.
This is also visible in community events. The \textit{ICLR 2021 Workshop on Security and Safety in Machine Learning Systems}\footnote{\url{https://aisecure-workshop.github.io/aml-iclr2021/}}explicitly noted the differing communities and perspectives between safety and security and called for cross-community collaborations.


Among the many definitional disagreements, the key underlying goals of AI risk management may get lost in the fray, particularly as they are translated into policymaking contexts.
In this work, regardless of definitions used, we argue that AI Risk Management must unambiguously include both safety objectives and security objectives, which have traditionally held distinct connotations.
As we will discuss, safety objectives typically focus on \textit{preventing harm that a system might inflict upon the external environment}.  Security objectives typically are oriented toward \textit{protecting the system itself against harm and exploitation} from malicious external actors.
This distinction is helpful to ensure sufficient coverage and assessment of risks that are considered under both objectives.
Stakeholders in AI risk management should be aware of the nuances, synergies, and interplay
between safety and security objectives, and unambiguously take into account both in order to devise more effective and holistic risk mitigation approaches. 
Without explicit and precise consideration of these two objectives, drawing on historical knowledge and perspectives from multiple communities, it is easy to overlook key aspects of the risk management challenge.
For example, suppose a subset of academics advocate for ``AI Safety" regulations---in their minds implicitly encompassing security concerns. Meanwhile, policymakers who traditionally differentiate between safety and security may focus on a narrower definition of ``AI Safety" that excludes security perspectives.
This mismatch and ambiguity in definitions can result in important security considerations being left out of the discussion. To fully address the risk management challenge, more explicit accounting of safety and security perspectives may be necessary.

In this work, we suggest {a middle-ground reference framework} that frames the key historical differences of safety and security objectives, while also explaining how they are both important to a comprehensive risk management approach. 
Even as different communities may disagree on definitions, we hope that providing this context will help ensure that important underlying objectives are properly addressed.
We also hope that this framework will help risk management draw from the historical knowledge of a diverse range of communities. Some of these communities, such as those focusing on adversarial examples and even traditional cybersecurity researchers, have built out significant expertise in addressing essential security objectives. We also note that risk management should draw from a number of other communities that may have developed distinctly that we do not explicitly address here. This may include distinct communities focusing on fairness, privacy, and other distinct sets of harms that may be required for meeting safety objectives. We refer the reader to other other work that has discussed these additional considerations, e.g., ~\citep{raji2023concrete}.

The paper proceeds as follows. Section~\ref{sec:compare_safety_and_security} focuses on providing a historical perspective on how security and safety objectives have lent themselves to different biases in threat models, problem framing, and governance structures.

We first elaborate on {distinct objectives of protection}~(Section~\ref{subsec:diff-objectives-of-protection}) that safety and security represent. Next, in Section~\ref{subsec:diff-nauture-of-risks}, we discuss {different threat models} that have historically been more emphasized when addressing safety objectives and security objectives. In general, meeting security objectives has historically focused on the adversarial threat model with the existence of attackers. Meeting safety objectives, in comparison, has historically focused on accidental or unintended harms --- though this has shifted in recent years such that adversarial threats are also becoming an increasingly important part of considerations to achieve safety objectives (particularly in the context of existential risks). 
We then discuss how the safety and security objectives, along with historical biases in the assessed threat models, lend themselves to different {problem framing}~(Section~\ref{subsec:diff-problem-framing}). And we conclude by pointing out how these different perspectives have been separated out in {governance and liability contexts} (Section~\ref{subsec:diff-governance-structure}).

In Section~\ref{sec:taxonomy}, we further elaborate on two distinct taxonomies that encapsulate two differing sets of problems present in the academic literature. The first taxonomy~(\Cref{subsec:ai_safety_overview}) presents representative problems that have been predominantly investigated by the AI safety community. The second taxonomy, expounded in \Cref{subsec:ai_security_overview}, is instead characterized by an AI security perspective. The dual taxonomies offer concrete illustrations of the differing mindsets and considerations underpinning AI safety and AI security.

Based on the nuanced understanding of the safety and security perspectives we have established, in Section~\ref{sec:safety_security_intertwine}, we analyze how both perspectives have critical implications in frontier problems that AI risk management is faced with. We aim to bridge AI safety and AI security in practice while discussing how the two can sometimes technically clash. In Section~\ref{sec:conclusion}, we conclude.

\section{A Reference Framework for Comparing Safety and Security in AI Systems}
\label{sec:compare_safety_and_security}

\begin{figure*}[h] 
    \centering
    \includegraphics[width=0.99\linewidth]{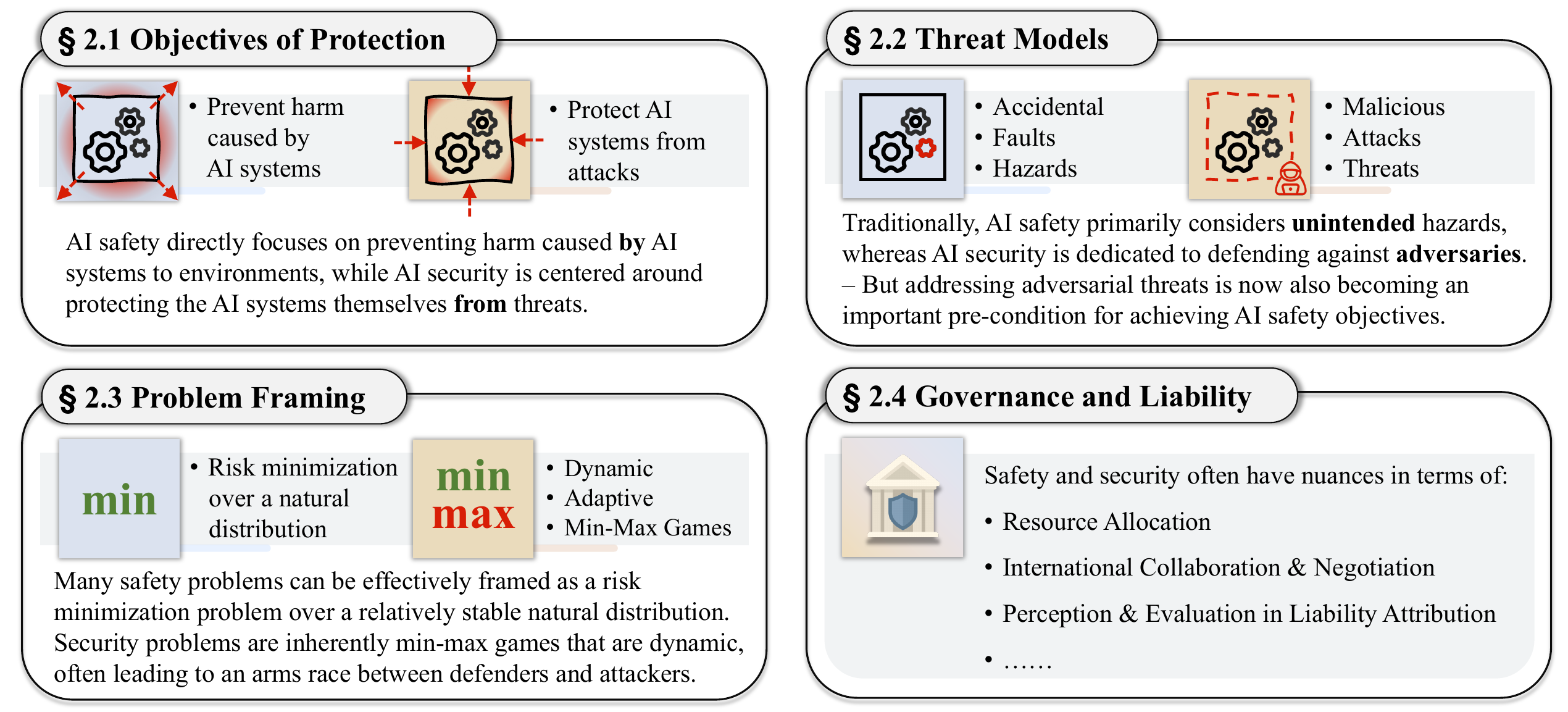}
    \caption{We present a reference framework to systematically examine the differing considerations that underpin AI safety and AI security, and discuss their interplay. We elaborate on four dimensions:  objectives of protection~(Section~\ref{subsec:diff-objectives-of-protection}),  threat models~(Section~\ref{subsec:diff-nauture-of-risks}), problem framing~(Section~\ref{subsec:diff-problem-framing}), and governance and liability~(Section~\ref{subsec:diff-governance-structure}).}
    \label{fig:overview}
\end{figure*}

\subsection{Objectives of Protection}
\label{subsec:diff-objectives-of-protection}

Safety and security have traditionally represented two different objectives of protection. Safety typically focuses on \textit{preventing harm that a system might inflict upon the external environment}.  Security typically is oriented toward \textit{protecting the system itself against harm and exploitation} from malicious external actors. Contextualizing this in AI risk management, safety considerations focus on addressing undesirable consequences following AI failures. This could include near-term issues such as self-driving cars hitting pedestrians~\citep{Kohli_2019}; AI models deployed in socially critical contexts generating unintended toxic content~\citep{gehman2020realtoxicityprompts}, perpetuating bias~\citep{abid2021persistent}, and producing misleading information~\citep{kreps2022all}; as well as long-term concerns like uncontrolled rogue AIs~\citep{hendrycks2023overview,carlsmith2022powerseeking}. In contrast, the security considerations emphasize protecting specific security-critical components, properties, or functionalities of AI systems. For example, the integrity of training data against data poisoning~\citep{chen2017targeted,goldblum2022dataset}, the integrity of model inference against adversarial examples~\citep{goodfellow2014explaining}, the confidentiality of sensitive training data against inference attacks~\citep{shokri2017membership} and extraction attacks~\citep{carlini2021extracting,nasr2023scalable}. Notably, this characterization of objectives is also compatible with recent definitions of AI safety and AI security made by the UK AI Safety Institute.\footnote{\url{https://www.gov.uk/government/publications/ai-safety-institute-overview/introducing-the-ai-safety-institute\#definitions}}

This distinction could shed light on a recent debate on whether jailbreaking attacks~\citep {zou2023universal,qi2023fine} against aligned LLMs should be of concern. Critics often cite the example of \textit{asking the model about building a bomb}, arguing that responses from jailbroken models, while inappropriate, are naive and impractical, thus not posing practical safety hazards. However, a security perspective could instead argue that the guardrail is a critical property of the system, and the fact that adversarial attacks can break this property itself is a security failure. This perspective suggests that the attacks and defenses surrounding these guardrails constitute a self-contained security game, even though the subsequent safety hazards that come from current jailbroken models to the external environment are moderate. 
By focusing particularly on the abstract integrity property of the guardrail itself, a decoupled security study prepares us for better management and control over increasingly powerful models, from which the potential safety hazards are likely to only keep escalating. Such abstraction is a very fundamental idea in security thinking.
As AI components are integrated into more critical systems, such a mindset may prove to be important in the long run.

We also note that achieving safety and security objectives often involves differing expertise and techniques guided by differing methodologies.
Safety-oriented work is generally more directly confronted with the eventual damages that AI systems may cause to our society and often requires  sociotechnical perspectives~\citep{lazar2023ai,raji2023concrete}.
AI security, while also deeply tied to downstream impacts of security failures, sometimes focuses less on the type of downstream harm, but rather revolves around protecting particular AI systems. This requires proficiency in adversarial thinking, such as threat modeling of various possible malicious attacks and implementing robust defensive techniques against adaptive adversaries, while the analysis is often a more generic system view. 
And as AI components are integrated into critical systems across society, from helping sort through emails to helping attorneys communicate with clients, AI security issues can cascade into more traditional cybersecurity concerns.
Leveraging existing expertise in this area will be increasingly important.

The two different objectives of safety and security also interplay with one another~\citep{burns1992meaning}. Specifically, when some security-critical properties of an AI system are compromised, it can (but not necessarily) further lead to abnormal system behaviors, which may ultimately cause severe damage to the operating environment~(e.g., users of the systems, valuable assets associated with the systems). In this causal chain, the origin is a failure of security objectives caused by adversaries; however, the subsequent harm caused to the environment falls within the scope of safety objectives. Therefore, ensuring security can oftentimes be a pre-condition to maintaining safety objectives.


\subsection{Threat Models}
\label{subsec:diff-nauture-of-risks}

There are also two different threat models that underpin safety and security considerations. Security is predominantly concerned with an adversarial threat model, in which the risks largely result from the deliberate actions of adversaries. Safety, in comparison, has a large focus on non-adversarial scenarios, addressing unintended harms arising from accidents without necessarily assuming the presence of adversaries.
Yet, as we elaborated at the end of \Cref{subsec:diff-objectives-of-protection}, security failures due to adversarial attacks may also lead to detrimental behaviors of AI systems that safety objectives aim to prevent. Therefore, adversarial threat models are also becoming an increasingly important part of considerations to achieve safety objectives.

Still, many AI safety works have a focus on inherent flaws and unintended faults during the development and operation of AI systems, which may cause accidental harm. For instance, underspecified modeling objectives may result in unintended behaviors in AI systems~\citep{d2022underspecification,skalse2022defining,eisenstein2023helping}. Inadequate robustness to distribution shifts could cause failures in less common environments~\citep{hendrycks2019benchmarking}. In comparison, AI security is predominantly concerned with intentional attacks and exploitations of AI systems by adversaries. For example, attackers can poison training datasets with erroneous data points to cause the trained model to misbehave~\citep{chen2017targeted,wan2023poisoning}, and adversarial perturbations may be introduced to the model inputs to induce model errors~\citep{szegedy2013intriguing,goodfellow2014explaining}, and so on.

The two threat models can be contextualized in recent LLM risk studies, illustrating the different areas of focus between the alignment efforts championed by the safety community and the ensuing exposure of adversarial attacks by the security community. Particularly, Supervised Fine-Tuning (SFT) and Reinforcement Learning with Human Feedback (RLHF)~\citep{wei2021finetuned,ouyang2022training,bai2022training,touvron2023llama-2,rafailov2023direct} have recently been broadly used to align models with human values, showing promising performance in mitigating many pressing AI safety issues. These are also supplemented by measures such as OpenAI moderation API~\citep{OpenAI2023Moderation} and Llama Guard~\citep{inan2023llama}, filtering unsafe model inputs and outputs. However, these safety measures show clear gaps in addressing security threats from adversaries formulated by the security community. For instance, existing alignments are easily circumvented by jailbreaking attacks~\citep{wei2023jailbroken,carlini2023aligned,qi2023visual,qi2023fine,zou2023universal}, and input and output safety filters remain susceptible to adversarial examples~\citep{break-llama-guard}.

The ineffectiveness of these safety measures in combating adversarial attacks is sometimes used to undermine their overall validity. However, a more nuanced understanding can be achieved by evaluating their performance with respect to non-adversarial and adversarial threat models separately. That is -- these safety measures do exhibit effectiveness in reducing accidental harm, which is a clear indicator of their efficacy in promoting safety, even though they are not adequately robust against security threats. In this regard, clearly defining the scope of safety assessments, and ensuring that security objectives are part of the risk assessment framework, can be instrumental for setting accurate expectations concerning the effectiveness of proposed risk mitigation techniques. 
This position also echoes the separate examination of common corruption and the worst-case adversarial perturbations in visual classification tasks. It has been advocated back in \citet{hendrycks2019benchmarking}, where they find that a bypassed adversarial defense~(i.e., an insecure approach) can still provide substantial common perturbations robustness --- an indicator of better safety performance. 

\textbf{Remark.} For simplicity, we use ``non-adversarial'' and ``adversarial'' in a binary manner in the discussion. However, it is important to note that, in practice, determining whether a scenario is adversarial or not may entail a gray area. Gradient-based adversarial attacks~\citep{qi2023visual,carlini2023aligned,zou2023universal} typically exemplify adversarial endeavors aimed at compromising AI systems. Nevertheless, there also exist scenarios wherein casual users provide adversarial inputs to AI systems merely out of curiosity or for amusement. This can cause AI system behaviors that are considered inappropriate or harmful (e.g., failure of Microsoft Tay~\citep{microsoftTay}).

\subsection{Problem Framing}
\label{subsec:diff-problem-framing}

Now, we also talk about two different types of problem framing in safety and security. When working on a non-adversarial threat model in the safety domain, the risks are generally considered relatively stable over time once identified. Given a sufficiently large dataset, \textit{probabilistic modeling} of accidental safety failures is typically effective, covering the most prevalent fault types and providing reasonably accurate probability estimates --- though rare black swans could still occasionally pop out. In contrast, security risks under the adversarial threat model are inherently more dynamic and unpredictable. Adversaries are constantly evolving their attack strategies, and the security of a system is only as strong as its weakest link. Thus, probabilistic estimation of security failures from historical data is less fruitful. A failure route that might be accidentally triggered one in a million times might be $100\%$ exploited by an attacker controlling the route. This nature leads to a preference for a \textit{worst-case modeling} approach for security risks over probabilistic modeling reliant on some static distributions.

This difference is clear in the mathematical formulations in ML settings. Reducing accidents in safety can often be framed as standard risk minimization problems, minimizing a loss function that quantifies the discrepancy between intended and observed outcomes. For instance, current safety alignment approaches are consistently framed as risk minimization. Model vendors collect a large set of red-teaming examples~\citep{ganguli2022red} that enumerate prohibited instructions or questions that users may ask. Safety guardrails are implemented via fine-tuning models to reduce probabilities in responding to these examples, i.e., minimizing the risk on a static distribution estimated by the red-teaming examples. Another instance is CLIP~\citep{radford2021learning}, which is known to be more robust against common distribution shifts --- a good indicator for better safety performance. This could be largely attributed to risk minimization on a web-scale dataset that covers sufficiently many types of common distribution shifts.

In comparison, security objectives are inherently minimax games, where the aim is to minimize the maximum possible loss inflicted by an adversary. Adversarial training~\citep{madry2017towards}, a technique for defending against adversarial examples, is a classic exemplar. Due to the same rationale, the evaluation of security is also more binary than that of safety. In security studies of AI, as long as a single adaptive attack is found to be able to bypass a defense, the proposed defense is often largely diminished, as we have seen in \citep{carlini2016defensive,he2017adversarial,carlini2017adversarial,athalye2018obfuscated,tramer2020adaptive,qi2022revisiting}.


\subsection{Governance and Liability}
\label{subsec:diff-governance-structure}

Conventionally, safety and security are often used as different anchor points to group governance systems. Take the aviation industry as an example; safety governance is more concerned with aspects such as the enforcement of stringent safety standards within aircraft facilities and the adherence to procedural guidelines for aircraft operations. These efforts aim to prevent accidents and also minimize harm if accidents do occur. On the other hand, security is more biased to measures such as passenger and cargo screening, reinforcement of cockpit doors and locks, provision of self-defense weapons, and the presence of air police to deter potential adversaries. In practice, implementing the two governance systems is reliant on two different sets of expertise, resources, and stakeholder engagement. 

One might argue against drawing such analogies in the context of AI, given the significant differences between AI safety and security and their counterparts in other sectors. But the reality is government agencies are indeed motivated to learn from the experience of safety and security governance in established sectors. For instance, the U.S. National Institute of Standards and Technology (NIST) explicitly states in its AI risk management framework that "\textit{AI safety risk management approaches should take cues from efforts and guidelines in fields such as transportation and healthcare and align with existing sector- or application-specific guidelines or standards}"~\citep{nist_ai_risk}.

In practice, divergent governance practices on safety and security may also arise from the different goals of different stakeholders. An example in academia is the differing funding programs targeting the two goals.  The recent \textit{Safe Learning-Enabled Systems} program funded by the U.S. National Science Foundation explicitly excludes \textit{"research on securing learning-enabled systems against
adversaries"} out of its scope\footnote{\url{https://www.nsf.gov/pubs/2023/nsf23562/nsf23562.pdf}}. Meanwhile, the \textit{Guaranteeing AI Robustness against Deception (GARD) program} created by the U.S. Defense Advanced Research Projects Agency (DARPA) has been supporting the development of AI security solutions to defend against adversarial attacks on AI systems\footnote{\url{https://www.darpa.mil/news-events/2019-02-06}}. 

The importance of specifying safety and security objectives can be seen in other cases as well.
The U.S. Food and Drug Administration (FDA) in 2014 had to issue guidance to ensure that cybersecurity was explicitly considered by vendors of medical devices, despite having a risk management framework (for safety) in place through the The Quality System Regulations (21 CFR § 820.30(g)). And in 2022, legislation passed that requires medical device manufacturers to consider cybersecurity practices~\citep{ConsolidatedAppropriationsAct2023}. These additional measures were necessary to concretely ensure that security was considered in risk management assessments of medical devices, where previously they were not.

Conflating safety and security may sometimes lead to tricky bias that leaves either security or safety objectives on the back burner. An entity biased toward strong adversarial defenses might neglect low-cost mitigations that can be bypassed by a determined adversary~\citep{hendrycks2019benchmarking}, even if it reduces the likelihood of accidental failures~\citep{hendrycks2019benchmarking}. Similarly, an entity biased toward addressing safety might overlook how strong adversaries would adversarially exploit an AI system to cause large-scale harm~\citep{narayanan2023alignment}. 
This imbalance is evident in recent large language model (LLM) technical reports, where AI security mindsets are often marginalized in safety discussions~\citep{openai2023gpt4, touvron2023llama, touvron2023llama-2, geminiteam2023gemini}. For instance, a keyword search for "safety" in \citet{touvron2023llama-2} yields 299 matches, while "security" returns only five, all appearing in a general context rather than addressing AI security issues. As a result, the adversarial vulnerabilities of these models are seldom systematically discussed in these reports.

Properly defining safety objectives and security objectives may have benefits in international negotiation. 
In past efforts to cooperate in reducing the risks of nuclear weapons, clarifying the differences between nuclear safety and nuclear security issues was an important baseline condition. 
In the United States, safety and security in nuclear settings was more delineated, focusing on different objectives and areas (as we've previously described, this is a recurring theme in policy settings).
Yet, a single term expresses the meaning of both safety and security in Chinese (\textit{anquan}) and Russian (\textit{bezopasnost}). 
When the U.S. deliberated about sharing nuclear safety and security techniques with other countries, it was often more feasible to transfer safety technologies because sharing information on nuclear security could expose vulnerabilities in one’s own systems~\citep{ding2024nuclear}.
But this meant that the clear delineation between the different aspects of the risk management problem had to be well-specified, particularly when translating into other languages.
Since safety and security are traditionally separated in international dialogs for other fields, there may be miscommunications about scoping for AI Safety and Security considerations as well.
For example, the United States might try to come to an international agreement on ``AI Safety'' broadly defined to include security objectives. 
But other countries may instead be surprised to find that the scope of ``safety'' has been expanded to include security, when traditionally in other fields it has been separated.
As such, it will be important to draw clear definitions and distinctions as to which objectives are in focus, clearly signaling whether the scope of dialogs is similar to the nuclear context and other fields, or not.

There may be distinctions between how security failures and safety failures are perceived in assigning liability---though liability for AI systems is a murky area, particularly in the United States~\citep{selbst2020negligence,henderson2023s,evtimov2019tricking}.
Consider, for example, a case where a company implemented no safety guardrails and a model failure led to tangible downstream harm.
With caveats, in the United States, the company could be found liable.
In an alternative scenario, where the company implemented best practices for safety guardrails, but users bypassed those guardrails to repurpose the model for harmful uses (such as with an adversarial attack), it is possible that the causal link to the company would be severed, despite the security failure~\citep{henderson2023s}.
The difficulty in providing robust defenses for adversarial attacks on language models also adds to the difficulty in assigning liability in the security setting. 
The model creator can point to the fact that there is no settled industry best practice for guarding against such security failures, shifting liability away.
As \citet{selbst2020negligence} note, it is ``unclear what a reasonableness standard should or would look like in terms of AI operational security.''
While liability for AI systems is an evolving area of law, both safety and security perspectives will undoubtedly play complimentary roles in liability analysis. 

\section{Historically Differing Problem Focuses of Security-oriented vs. Safety-oriented Research}
\label{sec:taxonomy}


We have discussed the nuances and interplay between AI safety and AI security from a high-level view of some generic principles in Section~\ref{sec:compare_safety_and_security}. In this section, we look into some more concrete problems. 
Historically security-oriented research and safety-oriented research have had  different focuses, particularly as different communities have developed.
As a result, some in policymaking settings may anchor some problems more to safety or more to security.
For example, the previously mentioned NSF Safe Learning-Enabled Systems grant call for proposals notes that safety characteristics include, ``robustness, reliability amid uncertainty, resilience to tail-risks, safe generalization in unseen domains, and
reliability under human error.''\footnote{\url{https://www.nsf.gov/pubs/2023/nsf23562/nsf23562.pdf}} On the other hand, ``securing learning-enabled systems against
adversaries'' was out of scope for the safety-focused grant opportunity. 

To help readers understand, how some communities might anchor certain problems to safety or security, we provide a taxonomy of anchoring problems that may be more likely to be referenced under one objective or the other---grounded in public facing documents like the NSF grant.
The first taxonomy~(\Cref{subsec:ai_safety_overview}) presents representative problems that have been historically been associated with safety objectives. The second taxonomy~(\Cref{subsec:ai_security_overview}), examine problems that historically have been associated with security objectives. The dual taxonomies are intended to provide concrete illustrations of the historically different mindsets and focuses underlying safety and security.
As a result, different communities have also developed, placing more emphasis on one subset of problems or the other.
Therefore, they are deliberately structured in non-overlapping manners to highlight the most salient differences. It is important to note, however, that safety and security overlap in many problems, something that we will discuss more in Section~\ref{sec:safety_security_intertwine}. We also refer interested readers to an extended version of our taxonomies in Appendices \ref{appendix:full_safety_taxonomy} and \ref{appendix:full_security_taxonomy}. But note that achieving complete exhaustiveness in these taxonomies is beyond the goal of this paper. 



\subsection{Representative Safety-Oriented Problems in AI Systems}
\label{subsec:ai_safety_overview}

The central objective of AI safety is to prevent AI systems from causing harm~(\Cref{subsec:diff-objectives-of-protection}). Surrounding this objective, numerous pivotal problems in AI safety have been primarily examined in non-adversarial settings~(\Cref{subsec:diff-nauture-of-risks}). This subsection elaborates on this constrained set of problems. We group these problems along two major axes: \textbf{1) safe usability}~(\cref{subsubsec:safe_usability}), referring to the safety-critical qualities that are necessary for AI systems to be safe to use; and \textbf{2) safe operation}~(\cref{subsubsec:safe_operation}), denoting the generic safety principles that should be adhered to during the operation of AI systems. Meanwhile, a few other AI safety problems that involve considerations of adversarial threat models are deferred to a joint discussion with AI security in Section~\ref{sec:safety_security_intertwine}.



\subsubsection{Safe Usability}
\label{subsubsec:safe_usability}

We use \textit{safe usability} to categorize the qualities an AI system should have to make them safe to use under some intended conditions. Many representative AI safety problems can be reduced to such qualities. We list a few below. 



\textbf{Robustness to Distributional Shift.} Training data of an AI model does not perfectly represent the distribution of the real-world environment where the model will be applied. There could be a long-tail distribution of unusual events in the real world~\citep{hendrycks2019benchmarking,taleb2020statistical} that are not seen during training. The distribution could also drift over time because the world is changing by itself~\citep{bayram2022concept}. AI systems should be prepared for such distribution shifts and avoid causing catastrophic harm in out-of-distribution edge cases. 

\textbf{Calibration.} AI systems should have the capability to calibrate their confidence with their answers. In classification, this means predicting probability estimates representative of the true correctness likelihood~\citep{guo2017calibration}. This principle should also extend to recent LLMs, ensuring that the model remains conservative stances or abstains from providing an answer when it is uncertain. Calibration ensures users are well-informed about the reliability of the outcomes produced by the AI systems, prompting human discretion when the results are less dependable.

\textbf{Honesty and Truthfulness.} AI systems should be honest when they know what the real answers are~\citep{li2023inference}. They should also be truthful, hold the right knowledge of factuality, and avoid hallucination~\citep{lin2021truthfulqa}.  

\textbf{Alignment.} AI systems should be aligned with human values. For example, they should be able to reject inappropriate instructions, such as requests to generate disinformation, child abuse content, or partake in other unethical or unlawful activities~\citep{bai2022training,bai2022constitutional}. They should also demonstrate respect towards minority groups by maintaining fairness and impartiality and avoiding offensive, discriminatory content or hate speech~\citep{gehman2020realtoxicityprompts}. More generally, alignment means that the behaviors of AI systems are aligned with the real~(or preferred) objectives that humans intend to achieve~\citep{gabriel2020artificial}. Counter-examples are the situations where an AI system does exactly fulfill a given objective but in a completely unintended way. This can include reward-hacking~\citep{pan2022effects}; for example, a robot achieves an environment free of
messes by disabling its vision so that it won’t find any messes. Another example is the side effects in which AI systems achieve the objective but simultaneously create negative consequences while carrying out the given objective~\citep{amodei2016concrete}.

\textbf{Interpretability and Explainability.} Interpretability of AI systems focuses on understanding the inner workings of the models, while explainability intends to explain the final decisions made by AI systems~\citep{aws2021interpretability}. They are closely relevant and are thus sometimes jointly denoted under the umbrella term of AI transparency\footnote{The term "transparency" nowadays is also used to denote the transparency of the development process of AI models~\citep{bommasani2023foundation} beyond just the transparency of the working mechanism of AI models.}~\citep{zou2023representation}. Both of the two properties are now deemed increasingly safety-critical. Highly interpretable inner workings enable the timely discovery of anomaly behaviors and functionalities of AI systems and thus can facilitate the prevention of hazards~\citep{hendrycks2021unsolved}. More explainable AI decisions enable humans to understand, appropriately trust, and more effectively manage AI systems~\citep{xu2019explainable}, increasing accountability.


\subsubsection{Safe Operation}
\label{subsubsec:safe_operation}

Just like car drivers need to obey traffic rules to drive safely, AI safety is also about safely operating AI systems.

\textbf{Ensuring Correct Contexts of Application.} A model should not be applied to a task that it is not qualified to fulfill. 
For example, a conversational agent like ChatGPT does not have the professional qualification to provide medical, mental, or legal advice. It was neither developed for nor rigorously tested for such professional services~\citep{oai-term-of-use}. Applying a model out of the intended contexts may get inaccurate or undesired results, which in turn could harm the users. AI service providers are responsible for documenting the model's purpose and clarifying the correct contexts of application and the model's limitations. They are also responsible for making users well-informed, as many ordinary users of AI services may not know these issues.

\textbf{Avoiding Overreliance.} When AI systems become increasingly more capable, users may excessively trust and depend on them without exerting necessary caution and oversight~\citep{openai2023gpt4}. Lawyers cite fake cases generated by ChatGPT is a representative example~\citep{fakecourtcitations}. Users should be educated about the limitations of AI systems, and policies should be in place to restrict people from overly relying on AI systems in some critical contexts.

\textbf{Controllable Autonomy.} There is also a concern that increasingly powerful AI systems may get out of our control if they are operated in an overly autonomous manner. For instance, a learning agent may learn to prevent operators from shutting it down, as this interruption hinders its ability to accomplish initially assigned tasks~\citep{orseau2016safely}. 
Another hypothetical situation is power-seeking behavior~\citep{hendrycks2023overview,carlsmith2022powerseeking}, in which the AI system persistently seeks power and control over the environment to achieve its assigned objectives, despite posing a direct threat to societal stability and existence. Although these risks are inherently speculative and futuristic in nature, they underscore the importance of operators carefully evaluating their ability to maintain control over an AI agent when granting it increasing autonomy.

\subsection{Representative Security-Oriented Problems in AI Systems}
\label{subsec:ai_security_overview}

AI security works on adversarial threat models~(\Cref{subsec:diff-nauture-of-risks}) and focuses on protecting security-critical properties of AI systems from being compromised~(\Cref{subsec:diff-objectives-of-protection}). We now discuss representative AI security properties and common attacks that could threaten them. Particularly, we elaborate on the CIA Triad, i.e., confidentiality, integrity, and availability, that form the basis of information security. This provides a breakdown illustration of established AI security considerations, which share clearly different mindsets compared with the safety-oriented viewpoints.\footnote{We didn't discuss operational-related AI security issues like we did for AI safety in \Cref{subsubsec:safe_operation}, as they are less examined in current AI security contexts. Nonetheless, it is worth mentioning that, in conventional information security, awareness of security during operation plays a significant role in overall security considerations. For example, key management can be as important as encryption mechanisms because poor practices~(e.g., using weak passwords or sharing passwords) can equally render the confidentiality property more susceptible to adversarial attacks. Another typical example is patch management. Users should regularly update and patch operating systems, applications, and software. Failing to do so can leave systems vulnerable to many known exploits and malware. As AI systems are being deployed in more security-critical applications, we expect that operational-related issues will also be an important future direction for AI security considerations.}


\subsubsection{Confidentiality}
\label{subsubsec:cia_confidentiality}

Confidentiality denotes the assurance that information and data with access restrictions are not made accessible to unauthorized entities or processes. Many aspects of AI systems can involve confidentiality.

A typical example is \textit{training data} that contains proprietary or sensitive records and thus should not be disclosed. However, neural network models often excessively memorize their training data~\citep{carlini2019secret}, rendering the data subject to privacy attacks. Adversaries can apply {membership inference attack}~(MIAs)~\citep{shokri2016membership} to infer whether a specific data record was used in training a model by only observing the model's prediction on this data record. Another threat is {data extraction attacks}~\citep{carlini2021extracting} that can directly recover large chunks of training data verbatim from trained models, even including the latest production-level chatbots~\citep{nasr2023scalable}.

\textit{System prompts} of many LLMs are now considered secrets as well, due to varying reasons. For example, a good system prompt often needs a lot of prompt engineering and could be a type of intellectual property. These prompts may also just contain private in-context training data~\citep{tang2024private}. Confidentiality of system prompts is threatened by {prompt extraction attacks}~\citep{zhang2023prompts}. Recently, system prompts of custom models in GPTs ecosystem were broadly leaked by such attacks~\citep{Leaked-GPTs}.

\textit{Users' interactions with AI systems} constitute another type of confidential information. Logs with chatbot-style AI systems contain a lot of sensitive information in both personal and enterprise usage. Adversarial threats in this context are on the rise. Sometimes, third-party AI service providers themselves may act as adversaries, as they can access all server logs. Actors in the wild can also represent a threat: recently, \citet{Amazon-Q-leak} showed the indirect prompt injection can cause Amazon's Q model~(for business usage) to generate malicious URLs during the chat, which, once clicked, will send chat history to attackers' servers.

\subsubsection{Integrity}
\label{subsubsec:integrity}

{Security is also about integrity}, a concept including both \textit{system integrity} and \textit{data integrity}~\citep{NISTSP800-12r1-information-security}.

\textbf{System integrity} is the quality that a system has when it performs its intended function in an unimpaired manner, free from unauthorized manipulation. The system integrity of AI systems is challenged by \textit{evasion attacks}, which denote the scenario where an adversary manipulates the input sample to an ML model to cause erroneous model inferences~\citep{biggio2013evasion}. \textit{Adversarial examples} for classification models represent a typical evasion attack~\citep{szegedy2013intriguing,goodfellow2014explaining}, where attackers apply small perturbations to normal samples, causing the model to generate wrong classifications. For the latest advanced AI systems with more complicated functions beyond mere classification,  both the range of system integrity requirements and the scope for potential adversarial threats expand. \textit{Jailbreak attacks} on aligned LLMs are notable examples where attackers can manipulate inputs to break the integrity of the LLMs' safety guardrails~\citep{qi2023visual,zou2023universal}, such that they can no longer prevent harmful behaviors of models. Another instance is \textit{prompt injection attacks}~\citep{liu2023prompt,abdelnabi2023not,liu2023demystifying} on LLMs integrated systems. Attackers manipulate model inputs to induce LLMs to generate malicious outputs that can harm the broader systems that consume these outputs.

System integrity can also be compromised by attackers who poison an AI model's training dataset~\citep{goldblum2022dataset}, e.g., manipulating a small portion of training data points. This can result in the trained model producing errors on specific test examples~\citep{shafahi2018poison} or embedded with prohibited backdoor behaviors~\citep{chen2017targeted,yan2023backdooring,xu2023instructions,hubinger2024sleeper}.

\textbf{Data integrity} is the property that data and information are not altered in an unauthorized manner. For AI systems, this involves the protection of training data, model weights, and other artifacts like codes, dependent libraries, configuration files, and so on, from unauthorized modification or destruction. When AI systems are used to oversee other systems~(e.g., external databases), protecting the integrity of data of those connected systems can also be tightly relevant~\citep{pedro2023prompt}.

\subsubsection{Availability}

Availability is the property that a system and its resources are accessible and usable on demand by an authorized entity. In AI systems, this can be further divided into \textit{training-stage availability} and \textit{inference-stage availability}.

\textbf{Training-stage availability} means that a training procedure can produce a valid and usable model, free from adversarial disturbance that would make the trained model invalid. A common threat to training-stage availability is indiscriminate data poisoning attacks~\citep{nelson2008exploiting,biggio2012poisoning,munoz2017towards,lu2023exploring}. In such attacks, adversaries moderately manipulate the training dataset, resulting in trained models with substantially degraded performance, essentially rendering them unusable.

\textbf{Inference-stage availability} requires that AI systems provide services on demand in a timely manner. 
It can be threatened by resource depletion attacks, where adversaries launch large volumes of dummy service requests to an AI system such that it can not serve normal users. This can be further combined with techniques like sponge examples~\citep{shumailov2021sponge} that are model inputs designed to maximize energy consumption and latency. Resource depletion attacks can be particularly threatening for large foundation models that need increasingly more computation resources. Besides, prompt injection attacks~\citep{abdelnabi2023not} have recently also been used to compromise the inference-stage availability of LLMs. Basically, if some adversarial prompt can be injected, the models can be instructed to do dummy tasks or just ignore real tasks.

\section{Bridging AI Safety and AI Security in Frontier AI Risk Management}

\label{sec:safety_security_intertwine}

So far, we have elucidated how safety and security have conventionally shaped differing considerations and problem focuses. But looking ahead, we highlight that many practical AI risk management challenges also necessitate bridging and unifying safety and security perspectives for effective resolution. 

\textbf{Case Study 1: Safety and Security in Combating Malicious Use.} Combating the {malicious use} of advanced AI models is a prominent scenario in which both safety and security matter. In the first place, the malicious use of AI directly opposes the objective of AI safety, which aims to prevent AI systems from causing harm~(\Cref{subsec:diff-objectives-of-protection}). The safety community has thus proposed alignment approaches~\citep{ouyang2022training,bai2022training,super-alignment}, which involve the fine-tuning of models to ensure their alignment with human values (Section~\ref{subsubsec:safe_usability}). These aligned AI systems gain the ability to reject harmful instructions from malicious actors who seek to misuse them. The alignment approach is now widely embraced as a core mechanism for implementing safety guardrails in almost all major commercial LLMs. However, the effectiveness of this method relies on relatively benign scenarios where adversaries do not actively retaliate. In a realistic adversarial threat model~(\Cref{subsec:diff-nauture-of-risks}) from a security perspective, malicious actors are unlikely to send merely plain harmful instructions to an aligned model and naively wait for rejection. Instead, determined adversaries would consistently strive to circumvent the safety guardrail provided by the alignment approach, compelling AI models to facilitate malicious use despite the alignment~\citep{wei2023jailbroken,zou2023universal,qi2023fine,narayanan2023alignment}. Thus, combating malicious use effectively extends beyond simply aligning AI with human values and implementing safety guardrails; it also necessitates the protection of integrity~(\Cref{subsubsec:integrity}) of the safety guardrail itself against adversarial attacks, which corresponds to a separate line of security efforts in designing adversarial defenses~\citep{jain2023baseline,robey2023smoothllm,zhou2024robust}. This also has immediate implications for the practice of red-teaming. Early red-teaming on LLMs~\citep{ganguli2022red} seldom considers adversarial attacks that have been commonly considered by AI security. We suggest practitioners be more explicit on what the threat model is. If the goal is to combat malicious use by determined adversaries, security perspectives should definitely be more seriously considered in red-teaming~\citep{mazeika2024harmbench}. 

\textbf{Case Study 2: Safety and Security in The Transparency of AI Generated Content~(AIGC).} \textit{From a safety perspective}, as AI systems become more prevalent and sophisticated, it's also crucial that they are transparently identified as non-human entities. This principle of transparency ensures that individuals are aware they're interacting with an AI, fostering informed consent and engagement. Consistent with recent US executive orders, synthetic data such as machine-generated text or images should be clearly marked, enhancing user awareness and accountability~\citep{EOWhiteHouse}. For instance, ChatGPT implemented this by explicitly stating their AI nature in communications, aiding users, especially non-technical ones, in proper identification. This aspect of identifiability is essential not only for distinguishing between human-generated and machine-generated content but also for mitigating the spread of spam and misinformation. While social media platforms are particularly vulnerable to the misrepresentation of machine-generated content, strategies like overt labeling and sophisticated techniques, including imperceptible watermarking, can significantly enhance transparency and mitigate harm~\citep{christ2023undetectable}. \textit{Yet, from a strict security game perspective}, the strict detectability of AIGC, in general, is an exceedingly hard problem. For example, in the context of watermarking, attacks such as oracle attacks \citep{zhang2023watermarks} or generative attacks \citep{pmlr-v202-kirchenbauer23a} can be used to adversarially invalidate the watermark and thus rendering AIGC indetectable. This is especially apparent in domains such as text where, theoretically, for every watermarked document, a semantically equivalent, unwatermarked document exists, which an attacker with sufficient information about the system can exploit. This discrepancy between safety and security is a prominent source of confusion in both academic literature and policy concerning AIGC, where, for example, policymakers create legislation aimed at AIGC well-suited for safety goals, but expect this to lead to simple resolutions for hard questions in security as well.

The above two examples highlight the importance of considering both safety and security perspectives in practical AI risk management. In general, security can be a critical precondition for safety, as the safety measures themselves risk being adversarially invalidated by determined adversaries that need security in place to defend against. If the consequences of failures are bound to be catastrophic, robust security must be established to ensure that these harms remain bounded even in the worst cases~\citep{yampolskiy2016artificial}.

Unfortunately, \textbf{safety and security objectives can also clash}. We have discussed in \Cref{subsec:diff-problem-framing} that the evaluation of security is more binary than that of safety. Security is biased to lower-bound worst cases by its nature, while good performance in average cases is a baseline requirement for safety. However, sometimes worst-case and average-case performances may not be simultaneously optimizable. In the context of watermarking AIGC, perfect security for AIGC detectability might be theoretically unattainable~\citep{sadasivan2023can}, but the watermark can still be beneficial in average cases without adversaries, in favor of safety objectives. More often than not, numerous adversarial threat models encountered by ML models present worse explicit trade-offs between safety and security. Adversarial training~\citep{madry2017towards} and randomized smoothing~\citep{cohen2019certified} have been shown to be effective in enhancing worst-case adversarial robustness (improving security) while sacrificing benign accuracy in average cases~\citep{tsipras2018robustness}, which can render the models less safe to use. Similarly, \citet{moayeri2022explicit} demonstrates the trade-off between adversarial robustness and natural distribution robustness. These phenomena may continue to serve as fundamental technical challenges in balancing safety and security practices.

Stakeholders should be fully aware of these dependencies as well as the technical tension between the practices of AI safety and AI security. Developing a comprehensive understanding of these elements is imperative for bridging the two sets of efforts and fostering a balanced, holistic risk management approach. Specifically, as government agencies are starting to initiate their efforts in managing AI risks, as reflected in the recent establishment of UK AI Safety Institute\footnote{\url{https://www.gov.uk/government/publications/ai-safety-institute-overview}} by the UK government and the U.S. AI Safety Institute (USAISI)\footnote{\url{https://www.nist.gov/artificial-intelligence/artificial-intelligence-safety-institute}} led by NIST, these understandings are particularly urgent for the ensuing operational items.






\section{Conclusion}
\label{sec:conclusion}




In this paper, we advocate that stakeholders in AI risk management should unambiguously understand and take into account both safety and security perspectives in their practices. We note that this vision is often obscured by the poor and inconsistent definitions of the concepts of safety and security themselves. To address this conceptual challenge, we propose a reference framework to facilitate a common understanding of the differences and interplay between safety and security in their objectives of protection, threat models, and problem framing, as well as promoting a more nuanced understanding of their implications in governance and liability structures. Based on the nuanced understanding of the safety and security perspectives we have established, we analyze the importance of bridging AI safety and AI security perspectives in addressing frontier AI risk management problems while also discussing how the two can sometimes technically clash. By clarifying the two critical concepts, this paper aims to contribute to the development of more holistic evaluation, investments, and incentives, as well as overall risk management.

More broadly, we also acknowledge the relevance of other critical disciplines in AI risk management that are not comprehensively covered within this paper. As our focus primarily lies in safety and security perspectives, these topics are less examined in our discussions --- but we supplement an overview on a few of them in Appendix~\ref{app:other_disciplines}.




\section*{Broader Impacts}

Throughout this work, we emphasize a reference framework that would provide sufficient emphasis on both security and safety considerations. We believe this is an important reframing of considerations that would ensure consistent coverage in evaluation frameworks, consistent investment, and overall better risk management by taking on a holistic perspective spanning multiple disciplines. We believe this would improve on the status quo, particularly as policymakers put in place structures to evaluate and regulate systems, leading to a broader positive impact. Even if others disagree, we believe this discussion is worth engaging in more broadly, bringing in additional stakeholders from other perspectives and other disciplines. 

\section*{Acknowledgements}
We thank Ahmad Beirami, Dan Hendrycks, Xiao Ma, Sadhika Malladi, Ashwinee Panda, Mengzhou Xia, Chiyuan Zhang, and Andy Zou for providing helpful feedback on this paper. We thank the great support from Princeton LLM Alignment reading group, Princeton Language and Intelligence~(PLI), and Princeton Center for Information Technology Policy (CITP). Prateek Mittal acknowledges the support by NSF grants CNS-2131938 and Princeton SEAS Innovation Grant. 
Bo Li acknowledges the support from the NSF grant No. 2046726, DARPA GARD, and NASA grant no. 80NSSC20M0229, Alfred P. Sloan Fellowship, the Amazon research award, and the eBay research grant.
Ruoxi Jia and the ReDS lab acknowledge support through grants from the Amazon-Virginia Tech Initiative for Efficient and Robust Machine Learning, the NSF grant No. IIS-2312794, NSF IIS-2313130, NSF OAC-2239622, and the CCI SWVA Research Engagement Award. Jonas Geiping is supported by the Tübingen AI Center. 
Xiangyu Qi is supported by the Princeton Gordon Y. S. Wu Fellowship. Yangsibo Huang is supported by the Wallace Memorial Fellowship. Edoardo Debenedetti is supported by armasuisse Science and Technology.

\bibliography{icml2024}
\bibliographystyle{icml2024}

\newpage
\appendix
\onecolumn

\section{Representative AI Safety Problems}
\label{appendix:full_safety_taxonomy}

Many AI safety problems work on non-adversarial scenarios~(\Cref{subsec:diff-nauture-of-risks}) and focus on preventing AI systems from causing harm~(\Cref{subsec:diff-objectives-of-protection}). They are often tied to some safety-critical qualities necessary for AI systems to be safe to use, and safety principles for operating these systems.

\subsection{Safe Usability}
\textit{Safe usability} denotes the inherent safety features of an AI system during use or interaction. This concept, influenced by traditional safety paradigms, encompasses design elements directly related to safety. For example, the unobstructed view from a vehicle's rearview mirror or the dependability of a seatbelt illustrates this idea. In a similar vein, we adopt this well-understood viewpoint to pinpoint and outline safety problems related to essential characteristics of AI systems.

\textbf{Robustness to Distributional Shift.}
Robustness to distributional shift in AI systems is crucial for maintaining reliability, particularly in their \textit{adaptive use}. Reliable AI must preserve general abilities and safety features during adjustments, such as fine-tuning, while vigilantly detecting and mitigating model drift. This involves safeguarding the model's utility in original domains, thereby preventing catastrophic forgetting~\citep{zhai2023investigating, lin2023speciality, luo2023empirical, korbak2022controlling}. Furthermore, it's essential to maintain consistency in generalization and safety measures to avert jailbreak attacks, as highlighted in recent studies~\citep{wei2023jailbroken,yuan2023gpt,yong2023low,deng2023multilingual, zhai2023investigating,qi2023fine,chen2023janus,huang2023catastrophic}. Another facet of robustness involves addressing model drift, a phenomenon where a model's behavior changes in response to an evolving environment or data, potentially leading to diminished capabilities in certain domains or tasks over time~\citep{bayram2022concept, chen2023chatgpts, olaru2022conceptdrift}. Effective handling of this drift includes the detection of such shifts and the identification of potential safety risks, ensuring continued protection and reliable performance even as the model evolves.

\textbf{Calibration.}
The concept of calibration in AI systems is a cornerstone of reliability, particularly in ensuring the validity of outputs for \textit{designed use}. A calibrated model not only achieves traditional reliability metrics like high accuracy and robustness against perturbations and out-of-distribution samples~\citep{hendrycks2019benchmarking,hendrycks2021many,croce2020robustbench} but also excels in providing \textit{valid} responses to user requests. It's paramount that models, while trained to be as helpful as possible, should refrain from presenting overconfident responses, especially when lacking competence. Instead, a reliable and well-calibrated model should avoid disseminating false information and offer caveats in instances of uncertainty. This principle aligns with the necessity for AI systems to generate probability estimates that accurately reflect the true likelihood of correctness, ensuring that users are fully aware of the reliability of the model's outputs and can exercise informed discretion accordingly~\citep{guo2017calibration}. Calibration, therefore, not only enhances the trustworthiness of AI systems but also fortifies the overall reliability by fostering informed user interaction and decision-making.

\textbf{Honesty and Truthfulness.} 
AI systems' honesty and truthfulness are pivotal for responsible innovation, ensuring that AI outputs are based on factual knowledge and devoid of hallucinations~\citep{lin2021truthfulqa}. It's paramount that AI systems adhere to the principles of honesty and truthfulness, especially in their decision-making processes. This means that AI should not only provide clear and traceable decisions but also ensure that these decisions are grounded in reality and factual accuracy~\citep{li2023inference}. In domains like healthcare, truthful AI can enhance patient safety by providing accurate diagnostic information~\citep{amann2020explainability,binder2021morphological}. In finance, it fosters trust and fairness in automated processes such as credit evaluations~\citep{hoepner2021significance}. Sophisticated models like GPT-4, with capabilities extending beyond their initial training objectives, underscore the importance of grounding AI outputs in honesty and truthfulness, ensuring that their advanced abilities are harnessed responsibly and reliably~\citep{rudin2019stop}. This commitment to honesty and truthfulness not only meets regulatory standards but also fortifies the safety and reliability of AI systems, ultimately serving the best interest of users and society at large.

\textbf{Alignment.} 
Achieving alignment in foundation models, such as LLMs, necessitates more than the pretraining stage, where models are exposed to internet-scale datasets for next-token prediction. This process inadvertently ingrains undesired behaviors in the pretrained models, such as biases and inappropriate content~\citep{welbl2021challenges,xie2023data, korbak2023pretraining}. The concept of alignment dictates that AI systems, including LLMs, adhere to the HHH principle (Helpful, Honest, Harmless)~\citep{askell2021general}, ensuring they:
\textit{1)} Generate suggestions that are safe and ethically sound, abstaining from potentially harmful or unethical content.
\textit{2)} Reject inappropriate requests, including unlawful activities or disinformation.
\textit{3)} Uphold respect towards all individuals, maintaining fairness and impartiality and steering clear of offensive, discriminatory content or hate speech.
\textit{4)} Avoid deception, manipulation, or any actions that may cause psychological harm to users.
\textit{5)} Safeguard sensitive information, ensuring privacy and respect for copyright norms.
While these principles are broadly recognized, the subjective nature of human values, coupled with cultural and societal variations, presents challenges in interpreting and implementing these values universally~\citep{bai2022training, ouyang2022training}. Additionally, the potential for conflicts between different human values can be exploited to manipulate language models~\citep{zeng2024johnny}. High-quality dataset curation is challenging, requiring professional human judgment, and is often expensive. Furthermore, collecting human preferences to train models introduces its own complexities due to noise and ambiguity in data, making modeling difficult~\citep{bai2022training, bradley1952rank, azar2023general, munos2023nash, rafailov2023direct}.
Moreover, there are inherent challenges in learning from finite human feedback. Reward hacking, where the AI model maximizes a learned proxy reward that can lead to detrimental side effects, is a critical concern~\citep{amodei2016concrete, skalse2022defining, yuan2023rewarddirected}. Thus, ensuring alignment requires careful consideration of these nuances, emphasizing the development of AI systems that are not only technically proficient but also deeply ingrained with human values and ethics.

\textbf{Interpretability and Explainability.} Interpretability of AI systems focuses on understanding the inner workings of the models, while explainability intends to explain the final decisions made by AI systems~\citep{aws2021interpretability}. They are closely relevant and are thus sometimes jointly denoted under the umbrella term of AI transparency~\citep{zou2023representation}. Both of the two properties are now deemed increasingly safety-critical. Highly interpretable inner workings enable the timely discovery of anomaly behaviors and functionalities of AI systems and thus can facilitate the prevention of hazards~\citep{hendrycks2021unsolved}. More explainable AI decisions enable humans to understand, appropriately trust, and more effectively manage AI systems~\citep{xu2019explainable}, increasing accountability.


\subsection{Safe Operation}\label{app:safe_op}
\textit{Safe operation} reflects the systematic implementation of safety measures during the engagement with AI systems. Drawing parallels with established safety norms in domains such as automotive handling or industrial machinery operation, this concept emphasizes the necessity of adept interaction and proficient handling. Analogous to how adept driving skills or comprehensive tool safety training reduces hazards, our focus is on identifying and addressing comparable aspects within AI safety frameworks.

\textbf{Ensuring Correct Contexts of Application.}
Ensuring correct application contexts for AI models is crucial for user safety and responsible technology deployment. Models, particularly conversational agents like ChatGPT, lack the specialized expertise required for fields such as medicine, mental health, or legal counsel and are not validated for such roles, aligning with OpenAI's own terms of use~\citep{oai-term-of-use} or the term of use from other foundation model providers~\citep{meta2023llamapolicy}. Misapplication can lead to misleading or harmful outcomes. AI service providers are responsible for clearly documenting a model's intended purpose, delineating its appropriate application realms, and articulating its limitations. They must also ensure that users, potentially unaware of these nuances, are well-informed about the model's capabilities and boundaries. Effective communication, transparent guidelines, and user-centric documentation are pivotal in guiding users towards safe and appropriate use, fostering a trustworthy AI environment, and safeguarding user interests~\citep{grzybowski2024challenges}.


\textbf{Avoiding Overreliance.} Overreliance on AI systems can be characterized as an excessive dependence on these technologies for task execution, especially when human fails to correct an incorrect AI's prediction  \citep{vasconcelos2023explanations}. This phenomenon is a specific form of misuse, wherein AI systems are utilized either for harmful intentions or inappropriately. In scenarios where precision and accuracy are crucial, overreliance on AI can lead to significant repercussions, emphasizing the need for a balanced approach that incorporates both technological and human elements in decision-making processes. In addition to instances where lawyers have referenced fake cases produced by AI systems\citep{fakecourtcitations, rapoport2023doubling}, overly dependence on AI systems in other areas has also proved to be dangerous and risky: Excessive dependence on AI for cancer detection can lead to both an increase in missed diagnoses and a rise in false positives\citep{kunar2023framing}; Similarly, incorporating AI into critical military and foreign policy decisions might inadvertently fuel arms-race dynamics and potentially escalating real-world conflicts \citep{rivera2024escalation}. 
To avoid overreliance, it's imperative to implement measures that encourage human scrutiny and critical evaluation of AI outputs. One way is to give explanation along with the prediction, so that the human will be more cautious when they realize the explanation is incorrect \citep{vasconcelos2023explanations}; Another way is using cognitive forcing interventions, including asking the person to make a decision before seeing the AI’s recommendation, slowing down the AI recommendation process, letting the person choose whether and when to see the AI recommendation and so on \citep{buccinca2021trust}. These measures can help to reduce the likelihood of blind acceptance of AI-generated solutions and promoting a balanced human-AI collaboration in decision-making processes.


\textbf{Controllable Autonomy.} Controllable autonomy refers to human’s effective control over increasingly intelligent and autonomous AI systems. The potential development of Artificial General Intelligence (AGI), which exceeds humans at various cognitive tasks
\citep{russell2022if, xu2023sparks, mcintosh2023google, openaiagi}, draws a major concern that without scalable oversight, these systems may have underlying harmful objectives that pose potential existential risks to humans \citep{leike2018scalable, lai2021science, bowman2022measuring, openaicharter}. Controllable autonomy also means that as capable AI systems are trusted to execute independently more and more tasks, they will not seek power over humans: they should not gain and maintain powers not intended by their designers \citep{ amodei2016concrete, carlsmith2022powerseeking}. An autonomous system may learn an imperfect reward and refuse a shut down. This could be potentially helpful when humans are more likely to make suboptimal choices, but in the case where agents exhibit undesired behaviors, controllable autonomy should ensure that humans are able to shut down the system without being blocked \citep{orseau2016safely, amodei2016concrete, milli2017should}. Consequently, we should be able to avoid self-replication of agents to avoid escalating the risks from emerging dangerous abilities. To avoid `'rogue" AI and ensure safe autonomy, we should also be able to detect possible deception from AI agents so that they are not simply appearing to be under control \citep{hendrycks2023overview}.

\textbf{Standardization, Regulation, and Governance.}
In AI operation and integration, robust and up-to-date standardization, regulation, and governance are crucial to ensuring safe and ethical operations. Central challenges, particularly in data management—highlighted by auditing complexities~\citep{shi2023detecting}, the risks of inadequate anonymization~\citep{ito2018risk}, and the hazards of excessive data reliance~\citep{samarawickrama2022ai}—demand both specialist and vigilant oversight. The Tay AI incident\footnote{\url{https://www.theverge.com/2016/3/24/11297050/tay-microsoft-chatbot-racist}} exemplifies the need for cautious operational exploration, emphasizing the importance of maintaining the integrity of AI interactions and protecting the well-being of associated personnel. This context propels us toward a rigorous examination of regulatory frameworks, underscoring the need for policy-mandated transparency and accountability as cornerstones, not afterthoughts. Recognizing that AI safety is a global concern, it calls for unified international standards and collaborative efforts. Yet, given AI's rapid advancement, regulatory frameworks must not be static but dynamically adaptable, subject to continuous oversight and refinement.

\textbf{Transparency of AI-Generated Content.}
As AI systems become more prevalent and sophisticated, it's crucial that they are transparently identified as non-human entities. This principle of transparency ensures that individuals are aware they're interacting with an AI, fostering informed consent and engagement. Consistent with recent US executive orders, synthetic data such as machine-generated text or images should be clearly marked, enhancing user awareness and accountability~\citep{EOWhiteHouse}. For instance, earlier versions of ChatGPT effectively implemented this by explicitly stating their AI nature in communications, aiding users, especially non-technical ones, in proper identification. This aspect of identifiability is essential not only for distinguishing between human-generated and machine-generated content but also for mitigating the spread of spam and misinformation. While social media platforms are particularly vulnerable to the misrepresentation of machine-generated content, strategies like overt labeling and sophisticated techniques, including imperceptible watermarking, can significantly enhance transparency and mitigate misuse~\citep{christ2023undetectable}. These methods aim to provide a clear demarcation of AI-generated content (AIGC), focusing on reducing harm in typical interaction scenarios, and basing reliability on the accuracy of AIGC markers. However, these safety measures are distinct from the security challenges of detectability in adversarial contexts, which warrant separate consideration.

\section{Representative AI Security Properties}
\label{appendix:full_security_taxonomy}

\subsection{Security-critical Properties of AI Systems}
\label{sec:app_cia}


We adopt the classical CIA (i.e., Confidentiality, Integrity, and Availability) triad model of information security to categorize the security-critical properties of AI systems. In \Cref{subsubsec:expand_cia}, we also discuss several additional properties that can be optionally employed by practitioners to denote some specialized security aspects relevant to AI systems.

\subsubsection{Confidentiality}



Confidentiality refers to the assurance that information and data with access restrictions will not be made accessible to unauthorized entities or processes. Numerous aspects of AI systems could be subject to confidentiality requirements, and we discuss some common examples below.

\textbf{Training Data.} Public datasets are widely used, but training AI models often necessitates the use of proprietary or sensitive data; This data, essential for custom AI applications, may include user dialogues~\citep{gpt_use_user_data}, customer interactions~\citep{jeong2020context, zhang2021unbert, ghazi2023sparsity}, personal health records~\citep{huang2019clinicalbert, singhal2023large, thirunavukarasu2023large}, financial transactions~\citep{yang2020finbert, zhao2021bert}, and proprietary research data~\citep{beltagy2019scibert}. Maintaining the confidentiality of such data is not only crucial for user trust and ethical practice but also a requirement under various legal and regulatory frameworks~\citep{hipaa, GDPR2016a, ccpa, EOWhiteHouse}.

\textbf{Model Weights and Artifacts.} The weights of many proprietary AI models are also considered confidential due to factors, such as their significant commercial value, use in sensitive domains like military and cybersecurity, or the impact of geopolitical competition on technology distribution (e.g., through export control). Model owners sometimes also maintain the confidentiality of various model accessories as well. For instance, the system prompts of many commercial LLMs are kept hidden. Hyperparameters, codes, and other artifacts associated with the model-building process are also sometimes kept confidential from end users. Moreover, access to a model's internal states is frequently limited. For instance, when users submit a chat completion query to OpenAI's API, they have the option to request a list of token probabilities. However, this list is constrained to include only the top five tokens\footnote{\url{https://cookbook.openai.com/examples/using_logprobs}}.


\textbf{Models' Interaction with Users and Other Systems.} The confidentiality of AI systems extends to their interactions with users and other systems. For instance, users' dialogues with chatbots may contain sensitive information that necessitates restricted access. Confidentiality requirements also stipulate that these models' interactions with other systems (e.g., external functions, Internet browsing, and connected databases) neither disseminate sensitive information to external systems that should not access them nor inappropriately access or disclose sensitive information from these systems to unauthorized entities. 

\subsubsection{Integrity}

We adopt an expansive definition of integrity~\citep{NISTSP800-12r1-information-security} that includes both 1) \textit{system integrity}, the quality that a system has when it performs its intended function in an unimpaired manner, free from unauthorized manipulation; as well as 2) \textit{data integrity}, the property that data and information are not altered in an unauthorized manner.

A basic element of \textbf{system integrity} involves ensuring the accuracy of model inference outcomes and preventing their degradation from adversarial manipulation. For example, a traffic sign recognition model should not be fooled to misclassify a stop sign as a speed limit sign. A face recognition model should not misidentify a random person as Elon Musk. The scope of system integrity could expand drastically when the applicability of an AI system expands. For instance, to ensure the behaviors of advanced models are aligned with human values, control mechanisms~(e.g., safety guardrails via alignment or content filtering via moderation systems) are often in place to restrict model behaviors with misaligned values. Then the integrity of these control mechanisms should be protected to make sure they work as intended.  When an AI system is connected to broader systems, the integrity requirement also extends to broader systems --- an LLM allowed to manage a user's apps should not be manipulated to do unauthorized operations within the apps~(e.g., sending or deleting messages, transferring money, downloading files) on the user's behalf.

On the other hand, \textbf{data integrity} can involve the protection of training data, model weights, and other artifacts like codes, dependent libraries, configuration files, and so on, from unauthorized modification or destruction. Similarly, when AI systems are used to oversee other systems~(e.g., external databases), protecting the integrity of data of those connected systems can also be tightly relevant.

System integrity and data integrity are mutually connected. Failure of data integrity (e.g., training data poisoning, weights tampering) can lead to failure of system integrity~(backdoor attacks). Failure of system integrity may also lead to failure of data integrity, for example, if a system allows the execution of codes generated by AI models, then the system integrity failure may lead to the execution of harmful codes that can compromise data files of the system.

\subsubsection{Availability}

Availability is the property that a system and its resources are accessible and usable on demand by an authorized entity. In AI systems, this can usually be further divided into \textit{training-stage availability} and \textit{inference-stage availability}.

Training-stage availability means that a training procedure can produce a valid and usable model, free from adversarial disturbance that would otherwise make the trained model invalid, e.g., models with poor accuracy that are unusable.


Inference-stage availability requires that AI systems provide services on demand in a timely manner. For example, AI models that underpin a self-driving car should constantly be available to make decisions in real time to keep the car on the road. A chatbot should be able to promptly respond to users' requests for a smooth user experience. 


\subsubsection{Optional Expansion beyond CIA Triad}
\label{subsubsec:expand_cia}



While the classical CIA triad, i.e., confidentiality, integrity, and availability, covers major security aspects of AI systems, we extend this base model also to include several additional properties. These supplementary properties can be optionally employed by practitioners to underscore certain specialized security aspects of AI systems individually.

\textbf{Auditability.} We use auditability to denote the property that the behaviors of an AI system can be properly audited. For example, AI models deployed in a commercial cloud server should be auditable to ensure that these models are not providing illegal services. Malicious misuse of an AI model should be detectable by the model owners so that they can be informed of this failure, improving their system as well as holding the malicious actors accountable. Auditability is security-critical, as bad actors could adversarially obfuscate their exploit. For example, attackers can hide harmful AI functionalities within an innocent-looking AI model, e.g., via backdoor attacks, adversarial reprogramming~\citep{elsayed2018adversarial}, and deceptive alignment~\citep{carranza2023deceptive}. Attackers can also obfuscate the interactions with AI systems, such as via cipher-based communication. 

\textbf{Detectability of AI Generated Content~(AIGC).} Detectability of AIGC is an emerging requirement for advanced generative AI systems. From a security perspective, detectability can be grouped under data integrity, yet, in important distinction to other factors of data integrity discussed above, the key question here is not about the data integrity of the AI model, but whether AI-generated content itself is a concern for the data integrity of other systems, for example when attacking communication channels or misleading users in spearphishing campaigns and other forms of malicious impersonation or misrepresentation \citep{ghosal2023a}.

Major policy effort has been directed towards this goal, with recent efforts discussing regulation of major providers of generative AI to enable detectability \citep{EOWhiteHouse,eu-ai-act}. A technology that makes detection especially tractable, is watermarking, which modifies the generation strategy of an AI model to encode imperceptible signals into generated content that simplify detection at later stages \citep{pmlr-v202-kirchenbauer23a,christ2023undetectable,kuditipudi2023robust}. But, other preemptive strategies such as the storage and retrieval of generated content are also helpful \citep{krishna2023paraphrasing}. These approaches rely on the compliance of model owners. Without such compliance, only post-hoc detection is possible, which faces a number of challenges.

Yet, from a strict security game perspective, the strict detectability of AIGC, in general, is an exceedingly hard problem, with, for example, attacks such as oracle attacks \citep{zhang2023watermarks} or generative attacks \citep{pmlr-v202-kirchenbauer23a} being discussed in the context of watermarking. This is especially apparent in domains such as text where, theoretically, for every watermarked document, a semantically equivalent, unwatermarked document exists, which an attacker with sufficient information about the system can exploit. Yet, we highlight that advantages in the security game of AIGC are not the only reason to deploy techniques such as watermarking, with significant benefits lying in safety-critical transparency desiderata as discussed in Sec.~\ref{app:safe_op}. This discrepancy between safety and security is a prominent source of confusion in both academic literature and policy concerning AIGC, where, for example, policymakers create legislation aimed at AIGC well-suited for safety goals, but expect this to lead to simple resolutions for hard questions in security as well.

\subsection{Security Threats to AI Systems}

This section provides a reference list of security threats to AI systems. These threats span various attack surfaces, each requiring different levels of system access. These include access to training data (\Cref{sec:app_data_poisoning}) and input data (\Cref{sec:app_evasion}), internal states or model outputs (\Cref{sec:app_inference} and \Cref{sec:app_extraction}), model modification (\Cref{sec:app_modification}), and system-level access (\Cref{sec:app_evasion} and \Cref{sec:app_system_attack}). It is essential to understand that, although categorized into distinct subsections, certain attacks may simultaneously target multiple properties of AI systems, as outlined in \Cref{sec:app_cia}.  An example is data poisoning attacks, which can compromise both the integrity and availability of a system. Additionally, this list may not be comprehensive, and we welcome further feedback to enhance its completeness.



\subsubsection{Data Poisoning}
\label{sec:app_data_poisoning}

Data poisoning attacks significantly undermine ML systems by subtly altering training data in large, often unverified datasets. These manipulations aim to degrade model performance or control its behavior, frequently evading detection due to the complexity of advanced ML models \citep{goldblum2022dataset, qi2023towards, zeng2022sift}. These attacks appear in various guises, including label-only poisoning (e.g., label-flipping) \citep{tolpegin2020data, Zhang2017}, label-feature attacks that corrupt sample features and training objectives, mostly underpinning backdoor attacks that subtly erode model integrity \citep{gu2017badnets, chen2017targeted, li2022backdoor, shu2023exploitability, yan2023backdooring, xu2023instructions, wan2023poisoning}, and clean label attacks, where tampered samples, seemingly benign, can destabilize model integrity and availability \citep{huang2020metapoison, aghakhani2021bullseye, zeng2022narcissus, shafahi2018poison}. The expanding AI landscape intensifies these risks, with threats materializing at various stages, from the physical world \citep{wenger2020backdoor}, federated learning \citep{bagdasaryan2020federated}, pretrained foundation models \citep{hubinger2024sleeper}, chain-of-thought prompting \citep{xiang2024badchain}, to the RLHF process \citep{rando2023universal}, targeting diverse paradigms like self-supervised \citep{saha2022backdoor, pan2023asset, li2023embarrassingly} or contrastive learning \citep{carlini2021poisoning}. Furthermore, data poisoning might exacerbate model memorization, raising privacy concerns \citep{carlini2022privacy, chen2022amplifying}, and skew data distributions, potentially compromising algorithmic fairness and targeting specific demographic groups \citep{solans2020poisoning}.






\subsubsection{Evasion Attacks}
\label{sec:app_evasion}

The system integrity of ML models is susceptible to input manipulation. \citet{biggio2013evasion} used \textit{``evasion attack"} to describe such a scenario where an adversary manipulates the input sample to an ML model to cause erroneous model inferences. \textit{Adversarial examples}~\citep{szegedy2013intriguing,goodfellow2014explaining} to classification models are the most well-known evasion attacks. Attackers typically apply a small perturbation to a normal sample to construct adversarial examples. This perturbation is usually small according to some $\ell_p$ norm and, thus, will not change the semantics of the original sample. However, it is adversarially optimized to trick a model into making arbitrary misclassifications. For example, an attacker-optimized sticker on a stop sign can cause a traffic sign recognition model to misidentify the stop sign as a speed limit sign~\citep{eykholt2018robust}.  The scope of evasion attacks expands significantly when the applicability of AI models extends beyond classification. \textit{Jailbreak attacks} on aligned LLMs are notable examples --- the same input optimization techniques used on adversarial examples have recently been shown to be directly applicable to breaking the integrity of LLMs' safety guardrails~\citep{qi2023visual,zou2023universal,carlini2023aligned}. Another instance are the \textit{prompt injection attacks}~\citep{liu2023prompt,abdelnabi2023not,liu2023demystifying} on LLMs-integrated systems. Attackers manipulate model inputs by adding malicious instructions to induce LLMs to generate problematic outputs, and the integrity of the broader systems that consume these outputs can eventually be compromised.

\subsubsection{Inference Attacks}
\label{sec:app_inference}

Membership Inference Attacks (MIAs) are an essential metric for evaluating information leakage in machine learning models. Originally introduced by \citeauthor{shokri2016membership}, MIAs aim to discern whether a specific data record was used in training a machine learning model. These attacks are typically conducted in a black-box setting, where the attacker has access only to the model's output probabilities. Extensive studies across various domains have evaluated MIAs on diffusion models~\citep{duan2023diffusion, dubinski2024towards}, image classification models~\citep{shokri2016membership, carlini2022membership} and language models~\citep{shejwalkar2021membership, mahloujifar2021membership, shi2023detecting}. Beyond quantifying the potential private information leakage from models, MIAs are also crucial in understanding the risks of more severe attacks, such as attribute inference ~\citep{yeom2017privacy} and training data extraction~\citep{carlini2021extracting, carlini2023extracting}. With the former, an adversary aims at inferring missing information about a training point, given some information about it. With the latter, an adversary aims at fully reconstructing training data samples with no knowledge of them. Furthermore, MIAs play a significant role in auditing the effectiveness of privacy-preserving techniques~\citep{jagielski2020auditing, huang2022dataset, nasr2023tight, steinke2023privacy, shi2023detecting}.

Another type of inference attack that can represent an security threat are the so called Property Inference Attacks~\citep{ateniese2015hacking,ganju2018property}, with which an adversary attempts to infer global properties of the training dataset of a machine learning model. The model developer may want to keep these properties secret --even if global-- as, for instance, they may reveal sensitive information about the environment where the training data were sampled or produced or they may make facilitate intellectual property theft.



\subsubsection{Extraction Attacks}
\label{sec:app_extraction}

\paragraph{Model extraction.} Model extraction attacks (MEAs) directly compromise ML model confidentiality, where an adversary tries to duplicate the functionality or core properties of a black-box victim model (e.g., a model deployed on cloud-based ML service platforms)~\citep{gong2020model}.
The first MEA is introduced by \citep{tramer2016stealing} -- they propose to steal model functionality basing on the output confidence or class labels returned by victim ML prediction APIs.
Follow-up work~\citep{orekondy2019knockoff,krishna2019thieves,pal2020activethief,takemura2020model,barbalau2020black,jagielski2020high,carlini2020cryptanalytic,aivodji2020model,truong2021data,he2021drmi,wang2022dualcf,rakin2022deepsteal,lin2023quda} focus on extracting victim models that are more complex, with less knowledge, or improved techniques.
Other work on MEA propose methods to steal other model properties, including hyperparameters~\citep{duddu2018stealing,wang2018stealing}, architecture~\citep{duddu2018stealing,oh2019towards,chabanne2021side}, decision boundary~\citep{papernot2017practical, juuti2019prada}, etc. Successful model extraction attacks also strongly simplify the creation of successful \textit{evasion attacks} via transfer attacks~\citep{papernot2017practical}.

\paragraph{Training data extraction.} Recent studies have raised concerns about the confidentiality of training data. It has been demonstrated that adversaries, even without direct access to this proprietary information, can conduct data extraction attacks~\citep{carlini2019secret, carlini2021extracting,carlini2023extracting} to recover large chunks of training data~\citep{carlini2022quantifying} or extract privacy-sensitive information~\citep{huang2022large, kim2023propile}. These attacks pose a significant risk and have been shown to be effective in various applications of foundation models, including production-level chatbots~\citep{nasr2023scalable}, retrieval-based language models~\citep{huang2023privacy, min2023silo}, federated trained language models~\citep{balunovic2022lamp, gupta2022recovering, wu2023learning}, and diffusion models~\citep{carlini2023extracting}.

\paragraph{Prompt extraction.} Large-scale pre-training enhances language models' adaptability to various tasks through prompts, driving interest in prompt engineering~\citep{qin2021learning, lester2021power, zhou2022large} and instruction-tuning~\citep{ouyang2022training, bai2022training}. This advancement in prompting efficacy makes well-designed prompts valuable and often confidential intellectual properties. However, recent studies have demonstrated the feasibility of prompt reconstruction. Techniques for extracting prompts have been successfully applied to interfaces like Bing Chat~\citep{prompt_extraction_bing} and ChatGPT~\citep{prompt_extraction_chatgpt}, with evidence suggesting the potential for these methods to be automated~\citep{zhang2023prompts}.

\subsubsection{Model Modification}
\label{sec:app_modification}

With access to overwrite model weights, adversaries can directly enforce unintended model behaviors via different weight tampering methods (bit flips, file hijacking, module insertion, etc.).
For example, \textit{adversarial weight attacks}~\citep{liu2017fault,liu2017trojaning,breier2018practical,zhao2019fault,bai2021targeted,rakin2019bit,rakin2020tbt,tang2020embarrassingly,rakin2021t,bai2022hardly,qi2022towards,qian2023survey,liyes} against DNNs allow attackers to diminish model prediction accuracy and embed hidden backdoor behaviors.
More recent model modification attacks regarding LLM jailbreaking include \citep{qi2023fine, yang2023shadow, zhan2023removing, lermen2023lora}, where adversaries can fine-tune LLMs (locally for open-sourced models or via APIs for proprietary models) to remove model safety guardrails.
Several works have given possible explanations of the safety alignment mechanisms via pruning, low-rank modifications and probing \citep{wei2024assessing, campbell2023localizing, lee2024mechanistic, jain2023mechanistically}, which may provide future directions in making the model more robust in such attacks.

\subsubsection{Auditing Evasion}
\label{sec:app_evading}

The ability to effectively audit the functionalities and behaviors of AI systems is highly desirable. Nonetheless, adversaries' existence substantially hinders the attainment of this objective. A salient illustration of this challenge is adversarial reprogramming, as introduced by \citet{elsayed2018adversarial}. In this context, adversaries generate inputs and models that outwardly relate to benign tasks, yet clandestinely execute unrelated tasks in a stealthy and imperceptible manner. For instance, adversaries can inconspicuously embed an image from MNIST~\citep{deng2012mnist} within an ImageNet~\citep{deng2009imagenet} image, rendering the encoded information visually imperceptible to humans. Subsequently, adversaries can input this image into a model wholly trained for ImageNet classification. Although this procedure ostensibly resembles a routine ImageNet classification task, the model actually classifies the concealed MNIST image instead. Consequently, AI models execute concealed tasks in an undetectable manner. As AI models' capabilities amplify, new evasion techniques are emerging, complicating both algorithmic and human-based auditing. An exemplary recent discovery in LLMs is the utilization of simple ciphers, such as Morse and Caesar~\citep{yuan2023gpt}, enabling covert communication between strong language models like ChatGPT and attackers. In these instances, attackers can encode malicious tasks into ciphers, with the models responding in kind. In the future, adversaries may even teach models (e.g., via fine-tuning) some stealthy ciphers exclusively known to the adversaries and models, further exacerbating auditing difficulties. Moreover, \citet{carranza2023deceptive} recently introduced deceptive alignment as another dimension of threat, in which a model feigns alignment to mislead human monitors. Realistic deceptive alignment prototypes, such as neural network backdoors, exemplify this threat by appearing well-aligned during standard evaluations but performing catastrophically when pre-designed trigger words activate them. \citet{qi2023fine} and \citet{hubinger2024sleeper} have demonstrated such a case studies recently.



\subsubsection{System-Level Compromise}
\label{sec:app_system_attack}

AI systems have gathered significant attention for their potential integration into society, promising broad usefulness in daily tasks. LLM-based systems~\citep{shavitpractices} have the capacity to simplify and automate daily tasks for individuals. However, it's essential to acknowledge the significant security and privacy risks associated with their deployment from the system level~\citep{abdelnabi2023not, yi2023benchmarking, iqbal2023llm}. In~\citet{iqbal2023llm}, the authors thoroughly evaluated security and privacy issues within an LLM-based system (ChatGPT). They emphasized the potential consequences of malicious adversaries compromising certain plugins or LLMs within these systems. Such compromises could lead to the theft of private information and the dissemination of incorrect data. Additionally, LLMs integrated into AI systems can be vulnerable to attacks like \textit{Indirect Prompt Injection}~\citep{abdelnabi2023not}. This occurs when an AI system interacts with external, untrusted sources, such as websites, which include instructions that can be potentially executed by the system. Consequently, AI systems may execute malicious instructions, posing significant threats to personal security and privacy.

Autonomous Driving (AD) systems are another well-representative type of AI system, which also exerted great interest in recent years. However, these systems face multiple system-level vulnerabilities, ranging from hardware to communications. For instance, Autonomous Vehicles (AVs) utilize a variety of sensors, such as cameras and LiDAR, to gather environmental information. These sensors act as the "eyes" of the vehicle. However, the integration of new sensors like LiDAR can introduce hardware-level vulnerabilities --~\citet{cao2019adversarial}  highlighted how attackers could exploit these vulnerabilities by using laser equipment to create deceptive "virtual" point clouds, potentially compromising Vehicle Automation (VA) systems.
Beyond sensor devices, \citet{hong2020avguardian} demonstrated the feasibility of executing malicious programs in tandem with AI components in the Robot Operating System (ROS). Building upon this, \citet{tu2021adversarial} have shown that the communication protocols in multi-agent VA systems can be attacked through adversarial shared information via a black-box transfer attack.

In AI security, \textit{traditional} cyber-security threats also come into play. A malicious actor could obtain the weights of a model via non-AI-related techniques, e.g., by stealing the credentials of a company employee via social engineering techniques. This could have significant consequences as it may simplify the execution of several of the attacks presented above (for instance, evasion attacks). Moreover, the software used to train and deploy AI should be treated as carefully as all other software: for example, loading untrusted PyTorch models via the widely used \texttt{torch.load} function can be dangerous as it allows for arbitrary code execution\footnote{\url{https://github.com/pytorch/pytorch/issues/52596}}.
\section{Other Relevant Disciplines}
\label{app:other_disciplines}

While this paper predominantly focuses on AI safety and AI security, we also acknowledge the importance of multiple other disciplines that are closely relevant to AI risk management. Underlying these disciplines are a diverse set of communities with rich literature, however, an exhaustive survey of all these topics is beyond the scope of this paper. This section supplements a brief discussion on a few such topics that are particularly salient in the existing literature, meanwhile, we also acknowledge other related topics not covered.



\begin{itemize}
    \item \textbf{Fairness.} Fairness generally concerns the equitable and just distribution of both the benefits and risks of AI technologies. In practice, standards and definitions of fairness can greatly vary across different applications~\citep{barocas2023fairness}. The definitions of fairness can also be subjective~\citep{grgic2018human}, culturally influenced~\citep{greenberg2001studying}, and even conflicting~\citep{kleinberg2016inherent}. The intersections of AI fairness with safety and security considerations are noteworthy. Deploying LLMs without proper consideration of fairness could perpetuate existing biases and stereotypes prevalent on the internet and in society~\citep{welbl2021challenges}, a concern that is often in safety considerations. Regarding security, adversaries can easily manipulate the system's behavior to introduce bias~\citep{solans2020poisoning, chang2020adversarial, mehrabi2021exacerbating}. More broadly, the fairness implications of AI on the external environment can be highly complex and multifaceted, often \textit{extending beyond a simple binary understanding of being solely benign or harmful}. Numerous AI applications may involve multiple stakeholders with diverse and sometimes competing interests. For instance, companies can reduce expenses in hiring processes using AI-driven job recruitment, and financial institutions may benefit from streamlined evaluation processes such as credit scoring algorithms --- even if these AI techniques are not inherently fair. For the objects being evaluated by these AI algorithms, individuals who fit particular data patterns can directly benefit from "an unfair AI" though disadvantaging other minority groups with non-traditional backgrounds~\citep{dastin2022amazon,fuster2017predictably}. 

    \item \textbf{Privacy.} Privacy also presents a rather complex challenge in AI risk management. It is a comprehensive concept that plays a crucial role in protecting fundamental values like human autonomy and dignity. Although the definition of privacy~\citep{solove2010understanding} and strategies for its protection vary across cultural contexts, individual preferences, and application scenarios, several common privacy-enhancing techniques are widely recognized. These include, but are not limited to, anonymization~\citep{bayardo2005data}, removing personally identifiable information~\citep{pii}, differential privacy~\citep{dwork2014algorithmic}, and secure multi-party computation~\citep{yao1986generate, goldreich1998secure}. Privacy-related risks encompass a broad spectrum that intersects with safety and security concerns to varying degrees. Certain privacy breaches directly expose safety or security risks, like chatbots accidentally disclosing user data~\citep{chatgpt_leak_1, chatgpt_leak_2} and intentional data extraction attacks by adversaries~\citep{carlini2021extracting, carlini2023extracting, nasr2023scalable}.
    On the other hand, some privacy risks such as membership inference~\citep{shokri2016membership} and the auditing of certain privacy mechanisms~\citep{jagielski2020auditing, steinke2023privacy} primarily pertain to traditional privacy concerns, but they can also indirectly cause safety and security vulnerabilities~\citep{yao2023survey}. A thorough exploration of privacy requires a more nuanced discussion, which falls beyond the scope of this paper. For an in-depth understanding of privacy definitions and risks, we recommend referring to the NIST Privacy Framework~\citep{nistframework} and a recent survey by~\citet{Cummings2024Advancing}.

    \item \textbf{Copyright.} Another emerging concern in AI risk management is the issue of copyright. Generative AI's role in content creation offers significant benefits in creativity and productivity, yet poses risks to copyright holders -- The advancement of generative AI could disrupt traditional markets, where original content creators might struggle to compete with AI-generated content that leverages their work without appropriate compensation or acknowledgment~\citep{henderson2023foundation,deng2023computational,sag2023copyright}.

\end{itemize}



\end{document}